\documentclass[twocolumn,showpacs,aps,prl,superscriptaddress]{revtex4}

\usepackage{graphicx}
\usepackage{dcolumn}
\usepackage{amsmath}
\usepackage{xspace}

\input pubboard/babarsym.tex

\def\mX {\ensuremath{m_X}}

\def\kskpi {\ensuremath{\KS \Kpm \pimp}}
\def\kkpi {\ensuremath{K \Kb \pi}}
\def\hc {\ensuremath{h_c}}

\newcommand{\BABARPubYear}    {07}
\newcommand{\BABARPubNumber}  {001}

\newcommand{\SLACPubNumber} {12665}

\begin{document}

\begin{flushleft}
\babar-CONF-\BABARPubYear/\BABARPubNumber\\
SLAC-PUB-\SLACPubNumber\\
\end{flushleft}

\title{\Large \boldmath A study of \B-meson decays to $\etac K^*$ and $\etac\gamma K^{(*)}$}

\date{\today}
%
\author{B.~Aubert}
\author{M.~Bona}
\author{D.~Boutigny}
\author{Y.~Karyotakis}
\author{J.~P.~Lees}
\author{V.~Poireau}
\author{X.~Prudent}
\author{V.~Tisserand}
\author{A.~Zghiche}
\affiliation{Laboratoire de Physique des Particules, IN2P3/CNRS et Universit\'e de Savoie, F-74941 Annecy-Le-Vieux, France }
\author{J.~Garra~Tico}
\author{E.~Grauges}
\affiliation{Universitat de Barcelona, Facultat de Fisica, Departament ECM, E-08028 Barcelona, Spain }
% ...
\collaboration{The \babar\ Collaboration}
\noaffiliation

\begin{abstract}
We present preliminary results of a study of the two-body \B-meson decays to a charmonium state ($\ccbar$) and a $K^+$ or $K^{*0}(892)$ meson using a sample of about 349~\invfb\ of data collected with the \babar\ detector at the \pep2\ asymmetric-energy \B\ Factory at SLAC. Here $\ccbar$ indicates either the \etac\ state, reconstructed in the \kskpi\ and $K^+K^-\piz$ decay channels, or the \hc\ state, reconstructed in its decay to $\etac\gamma$. We measure $\BR(B^0\to\etac K^{*0})=(6.1\pm0.8_{\mathrm{stat}}\pm1.1_{\mathrm{syst}})\times 10^{-4}$, $\BR(B^+\to\hc K^+)\times \BR(\hc\to\etac\gamma)<5.2\times 10^{-5}$ and $\BR(B^0\to\hc K^{*0})\times \BR(\hc\to\etac\gamma)<2.41\times 10^{-4}$, at the $90\%$ C.L.
\vfill
\begin{center}
{\it Submitted to the 2007 Europhysics Conference on High Energy Physics, Manchester, England.}
\end{center}
\end{abstract}

\pacs{13.25.Gv, 13.25.Hw}

\maketitle

The \B\ decays to S-wave charmonium states, like \jpsi\ and \etac, have been observed to occur with large branching fractions (\BR) of the order $10^{-3}$~\cite{ref:pdg2006}. Experimental study of $B$ decays to singlet states of charmonium, such as \etac\ and \hc, is more complicated than the $B$ decays to triplet states, such as \jpsi, \psitwos\ or \chicone, because one cannot exploit the cleaner signature of final states including a lepton pair. 
In this document, we report measurements of the branching fraction for the following decay modes: $B^0\to\etac K^{*0}$, $B^0\to \hc \Kstarz$ and $B^+\to \hc K^+$~\cite{ref:cc_note}. We also reconstruct the $B^+\to\etac K^+$ decay to be used as a ``control sample''. The branching fraction of $B^0\to\etac K^{*0}$ is currently known with a $40\%$ uncertainty, $(1.2\pm0.5)\times 10^{-3}$~\cite{ref:etackstar_note}, while $B$ decays to \hc\ have never been observed. The Belle collaboration studied the decay $B^+\to\hc K^+$ with $\hc\to\etac\gamma$ and reported $\BR(B^+\to \etac\gamma K^+) < 3.8\times 10^{-5}$ at the $90\%$ C.L. for an invariant mass of the $\etac\gamma$ pair in the range [3.47,3.57]~\gevcc~\cite{ref:belle_exclhc}. No other $B^+$ or $B^0$ decay modes with \hc\ have been studied yet. The \hc\ meson has recently been discovered by the CLEO collaboration as a narrow peak in $\psi(2S)\to\etac\gamma\piz$ decays at a mass of $3524.4\pm0.7~\mevcc$~\cite{ref:CLEO_1P1}, and this observation was confirmed by the E835 collaboration~\cite{ref:E835_1P1}.

In the simplest approximation, $B$ decays to a charmonium state and a $K$ or $K^*$ meson  arise from the quark-level process \mbox{$b \to \ccbar s$} 
. The colorless current \mbox{$\cbar \gamma_\mu (1-\gamma_5)c$}, which can create the S-wave states like \etac\ and \jpsi, can also create the P-wave state \chicone. It cannot, however, create the $0^{++}$, $2^{++}$ and $1^{+-}$ states \chiczero, \chictwo\ and $h_c$. Therefore $B$ decays to any of these three states have to be ascribed to more complex mechanisms, such as the interaction of two color-octet currents~\cite{ref:NRQCD}. In this scenario, $B$ decays to \chiczero, \chictwo\ or $h_c$ are expected to occur as abundantly as those to \chicone. $B$ decays to $\chicone K^{(*)}$ have branching fractions between $3.2\times 10^{-4}$ and $4.9\times 10^{-4}$~\cite{ref:pdg2006}. The $B^+\to\chiczero K^+$ decay has indeed been observed with a branching fraction of $(1.4^{+0.23}_{-0.19})\times 10^{-4}$~\cite{ref:pdg2006}. However $B$ decays to $\chictwo K^{(*)}$ and $\hc K^{(*)}$ have not yet been observed and upper limits on their branching fractions slightly exceed $10^{-5}$~\cite{ref:pdg2006}.  

In this analysis we reconstruct the \etac\ in the $K_S^0(\to\pipi)K^\pm\pi^\mp$ and $K^+K^-\piz$ decay modes, the \hc\ in its decay to $\etac\gamma$, and the \Kstarz\ in the mode $\Kstarz\to K^+\pi^-$. The \kskpi\ and $K^+K^-\piz$ final states are manifestations of the same decay mode, $K\bar{K}\pi$: they are chosen because they are among the easiest \etac\ decay modes to reconstruct and have a rather large branching fraction, $\BR(\etac\to K\bar{K}\pi)=(7.0\pm1.2)\%$~\cite{ref:pdg2006}. The $\etac\gamma$ decay of the \hc\ is chosen because it is expected to comprise about half of the total \hc\ decay width~\cite{ref:NRQCD}. We measure ratios of branching fractions with respect to that of $B^+\to\etac K^+$, $(9.1\pm1.3)\times 10^{-4}$, to cancel the $17\%$ uncertainty on $\BR(\etac\to\kkpi)$. 

The data used in this analysis were collected with the \babar\ detector at the \pep2\ \epem\ storage rings, and correspond to about 349~\invfb\ of integrated luminosity collected at  the \FourS\ resonance, comprising $384$ million \BB\ pairs. The \babar\ detector is described elsewhere~\cite{ref:babar}. Momenta of charged particles are measured in a tracking system  consisting of a five-layer, double-sided silicon vertex tracker and a 40-layer drift chamber (DCH), both in a 1.5-T solenoidal magnetic field. Identification of charged particles is provided by measurements of the energy loss in the tracking devices and by a ring-imaging Cherenkov detector. Photons are detected by a CsI(Tl)  electromagnetic calorimeter (EMC).
The \babar\ detector Monte Carlo (MC) simulation based on GEANT4~\cite{ref:geant4} is used to determine selection criteria and efficiencies.

The event selection is optimized by maximizing the quantity $N_S/\sqrt{N_S+N_B}$, where $N_S$ ($N_B$) represents the number of signal (background) candidates surviving the selection. $N_S$ is estimated on samples of simulated events, while $N_B$ is extrapolated from regions far from the signals on data. Simulated signal events and data are normalized to each other using the available measurements for $B\to\etac$ decays and assuming $\BR=1\times 10^{-5}$ for $B$ decays to \hc.

We select events with \BB\ pairs by requiring at least four charged tracks, the ratio of the second to the zeroth order Fox-Wolfram moment~\cite{ref:FoxWolfram} to be less than 0.2, and the total energy of all the charged and neutral particles to be greater than 4.5~\gev. 

Charged pion and kaon candidates are reconstructed tracks having polar angles in the region $0.41 < \theta < 2.54 \rad$, at least 12 hits in the DCH, a transverse momentum with respect to the beam direction larger than 100 \mevc, and a distance of closest approach to the beam spot smaller than 1.5~cm in the plane transverse to the beam axis and 10~cm along the beam axis. 
A $K^{*0}$ candidate is formed from a pair of oppositely charged kaon and pion candidates originating from a common vertex and having an invariant mass within 60~\mevcc\ of the nominal $K^{*0}$ mass~\cite{ref:pdg2006}.

Photon candidates are energy deposits in the EMC in the polar angle range $0.32 < \theta < 2.44 \rad$ that are not associated with charged tracks, have energy greater than 100~\mev, and have a shower shape consistent with that of a photon. A \piz\to\gaga\ candidate is formed from a pair of photon candidates with invariant mass in the range [115,150]~\mevcc\ and energy greater than 400~\mev. The mass of such candidates is constrained to the nominal \piz\ mass~\cite{ref:pdg2006} when subsequently computing kinematic quantities.

A \KS\to\pipi\ candidate is formed from a pair of oppositely-charged tracks originating from a common vertex and having an invariant mass within 20~\mevcc\ of the $K^0$-meson mass. Its  measured decay-length significance is required to exceed three standard deviations ($\sigma$). The candidate is constrained to the nominal $K^0$ mass~\cite{ref:pdg2006}.

The $B^{+,0}\to\etac K^{+,*0}$ candidates are formed by pairing a \Kstarz\ or $K^+$ candidate, referred to as the primary kaon, and a \kskpi\ or $K^+K^-\piz$ combination with invariant mass in the range $[2.75,3.35]~\gevcc$. The mass range includes the \jpsi\ resonance. The $B^{+,0}\to\etac\gamma K^{+,*0}$ candidates are formed by combining a \Kstarz\ or $K^+$ candidate, a photon with energy exceeding 250~\mev, and  a \kskpi\ or $K^+K^-\piz$ combination with invariant mass consistent with the \etac\ mass. We perform a vertex fit to the \B\ candidates and require the probability of the $\chi^2$ of the fit to exceed 0.002. We define two kinematic variables: the beam-energy substitued mass, $\mes=\sqrt{E^{2}_{\mathrm{beam}}-p^{2}_B}$ and $\Delta E=E_B-E_{\mathrm{beam}}$, where $p_B$ ($E_B$) is the reconstructed \B\ momentum (energy) and $E_{\mathrm{beam}}$ the beam energy, in the $e^+e^-$ center-of-mass (c.m.) frame. $B$ candidates are retained if they have \mes\ greater than 5.2~\gevcc\ and $\Delta E$ within [-24,30], [-40,30], [-34,30], and [-40,30]~\mev\ for the $\kskpi K^{*0,+}$, $K^+K^-\piz K^{*0,+}$, $\kskpi\gamma K^{*0,+}$, and $K^+K^-\piz\gamma K^{*0,+}$ combinations, respectively. We also require the absolute value of the cosine of the polar angle of the \B\ candidate momentum vector in the \epem\ c.m. frame to be smaller than 0.9.

To suppress background, the $K^+\pi^-$, $K^+K^-$, $K^+\KS$ and $K^+\pi^-\pi^+$ combinations with invariant masses within 30~\mevcc\ of the $D^0$, $D_s$ and $D^+$ meson masses~\cite{ref:pdg2006}, respectively, are excluded to form $B$ candidates. We also remove $K^+K^-$ combinations containing a primary kaon where the invariant mass of the combination is within 30~\mevcc\ of the $\phi$ meson mass~\cite{ref:pdg2006}.

In events where more than one $B$ candidate survives the selection, the one with the smallest $|\Delta E|$ is retained. In cases of $B$ candidates composed by the same final state particles, thus having the same value of $|\Delta E|$, we retain the one for which the primary kaon  has the largest momentum in the \epem\ c.m. system.

The samples surviving the selection include a signal component, a combinatorial background component given by random combinations of tracks and neutral clusters both from \BB\ and continuum events, and a component due to $B$ decays with a similar final state as the signal. Such ``peaking backgrounds'' exhibit the same distribution as the signal in \mes\ and $\Delta E$, but their $\kkpi(\gamma)$ invariant-mass distribution ($m_X$) is different. The signal content on data is therefore obtained by means of a maximum likelihood fit to $m_X$ for all candidates having \mes\ in the signal region $[5.274,5.284]$~\gevcc, after subtracting the combinatorial background. The $m_X$ distribution for the combinatorial background events is obtained by extrapolating into the \mes\ signal region the $m_X$ distribution measured in the \mes\ sideband, defined by $\mes < 5.26~\gevcc$. The correlation between $m_X$ and \mes\ is found to be negligible in the relevant regions.  A binned fit is then performed on the \mes-sideband-subtracted \mX\ distribution. 

\begin{figure}[!tb]
  \begin{center}
    \includegraphics[scale=0.2]{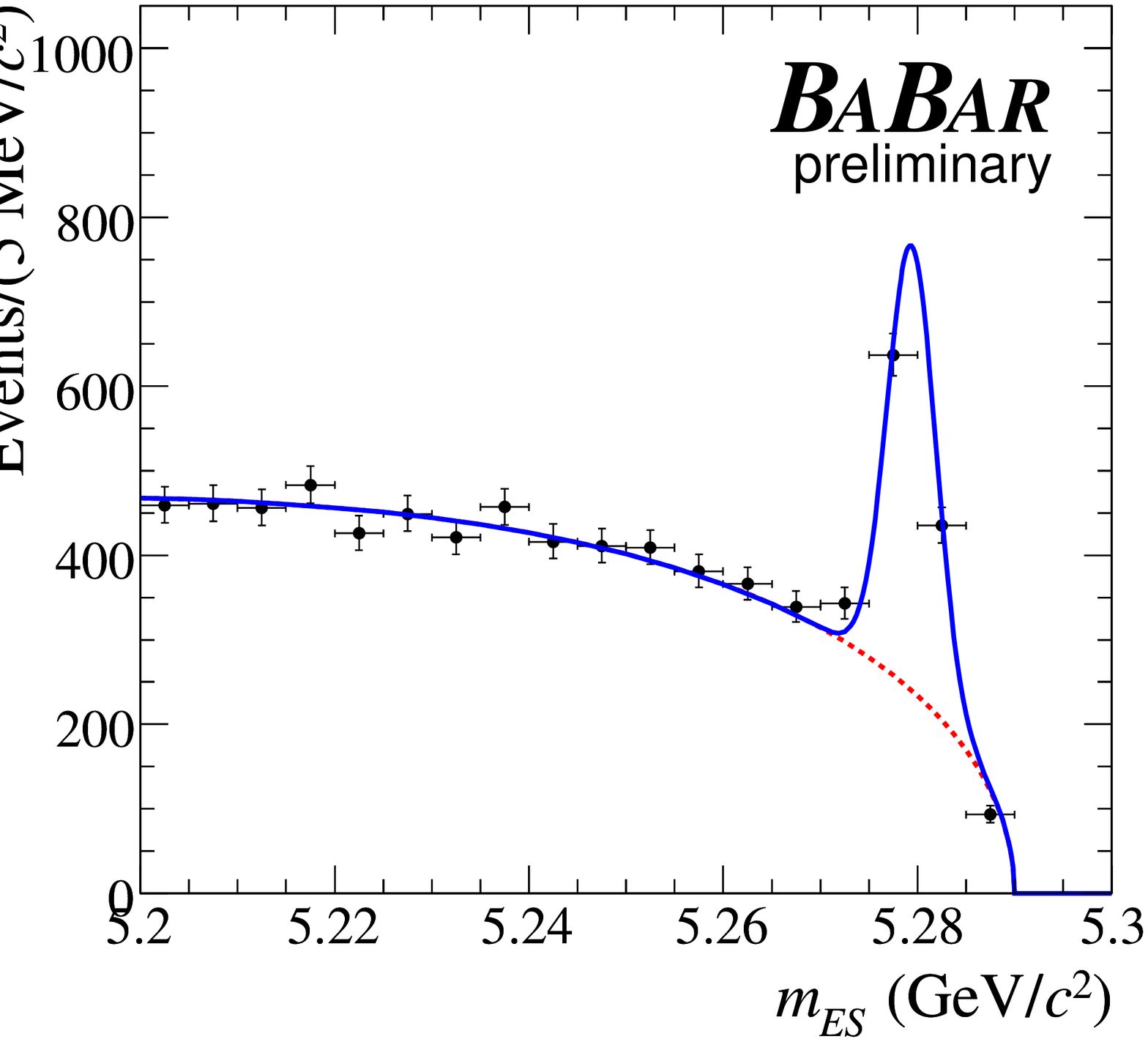}
    \includegraphics[scale=0.2]{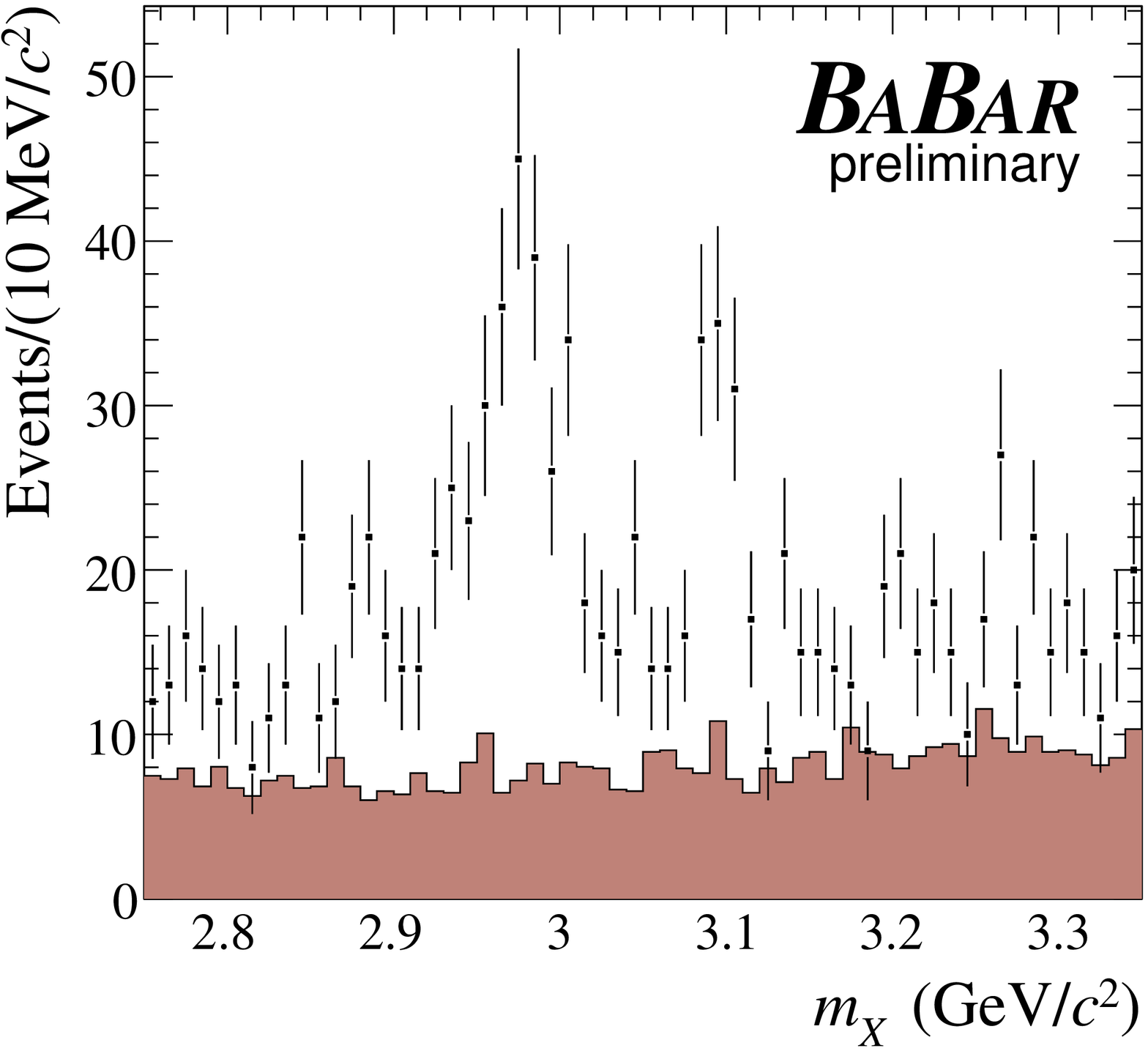}
    \includegraphics[scale=0.2]{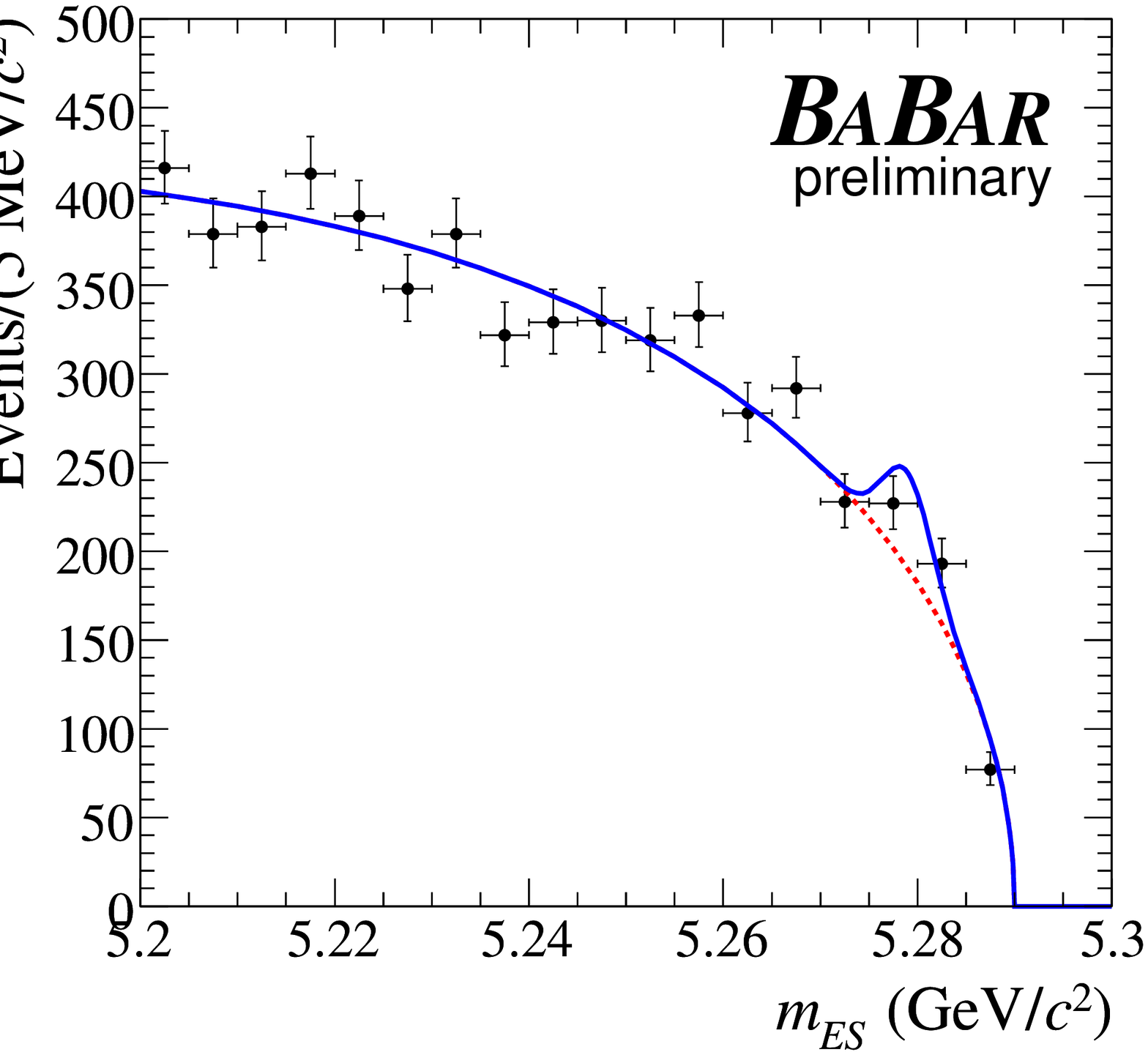}
    \includegraphics[scale=0.2]{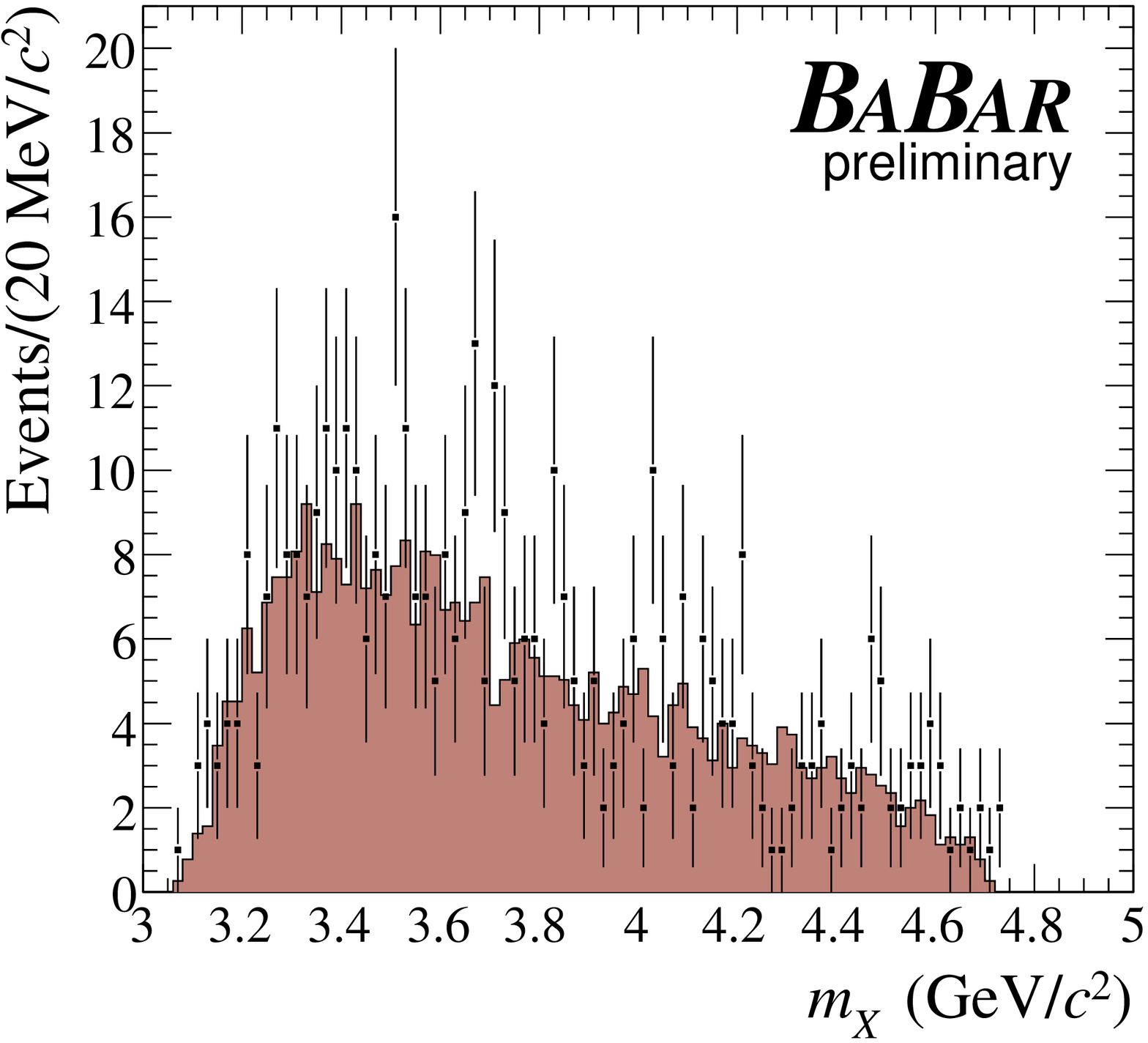}
    \includegraphics[scale=0.2]{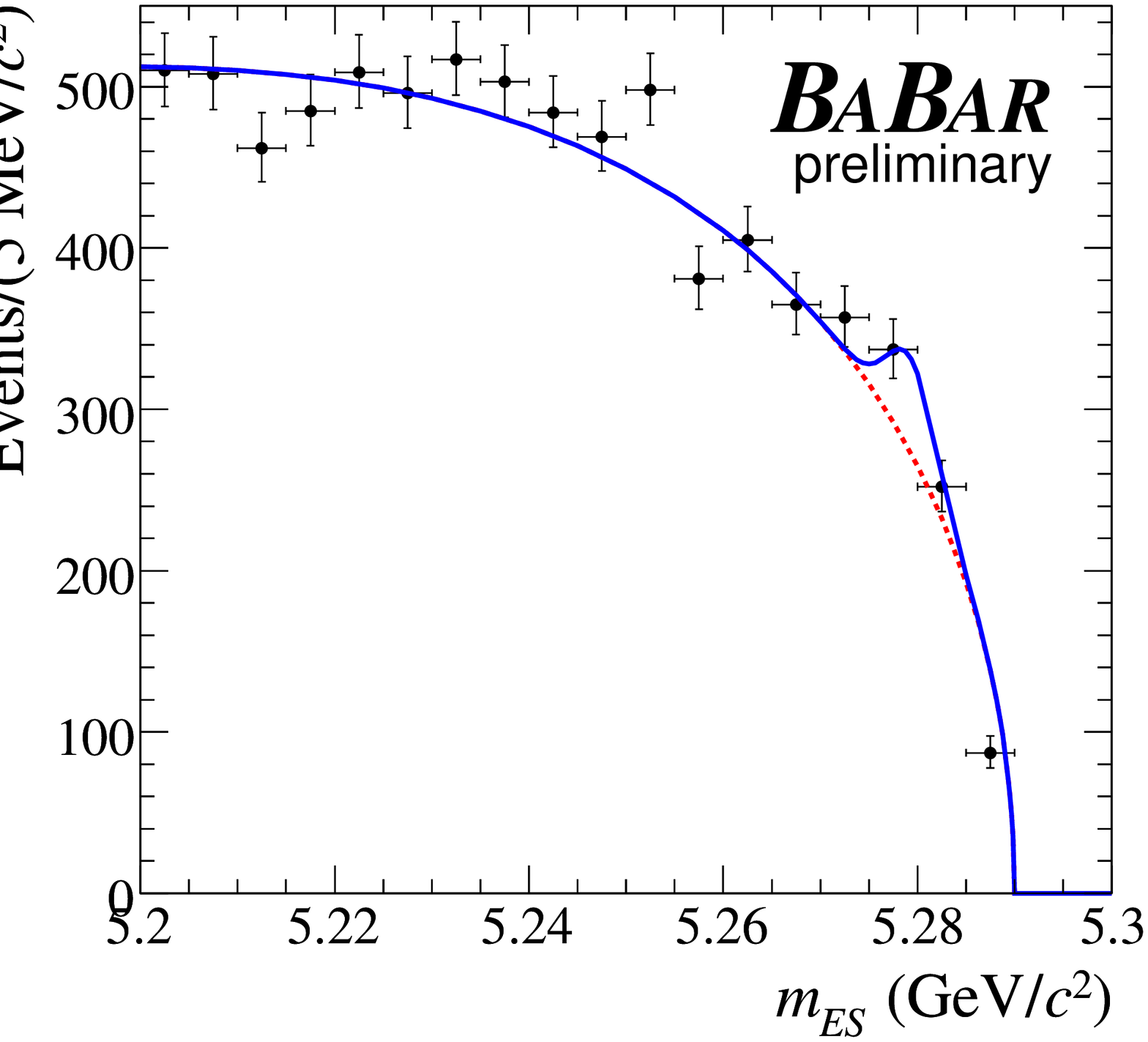}
    \includegraphics[scale=0.2]{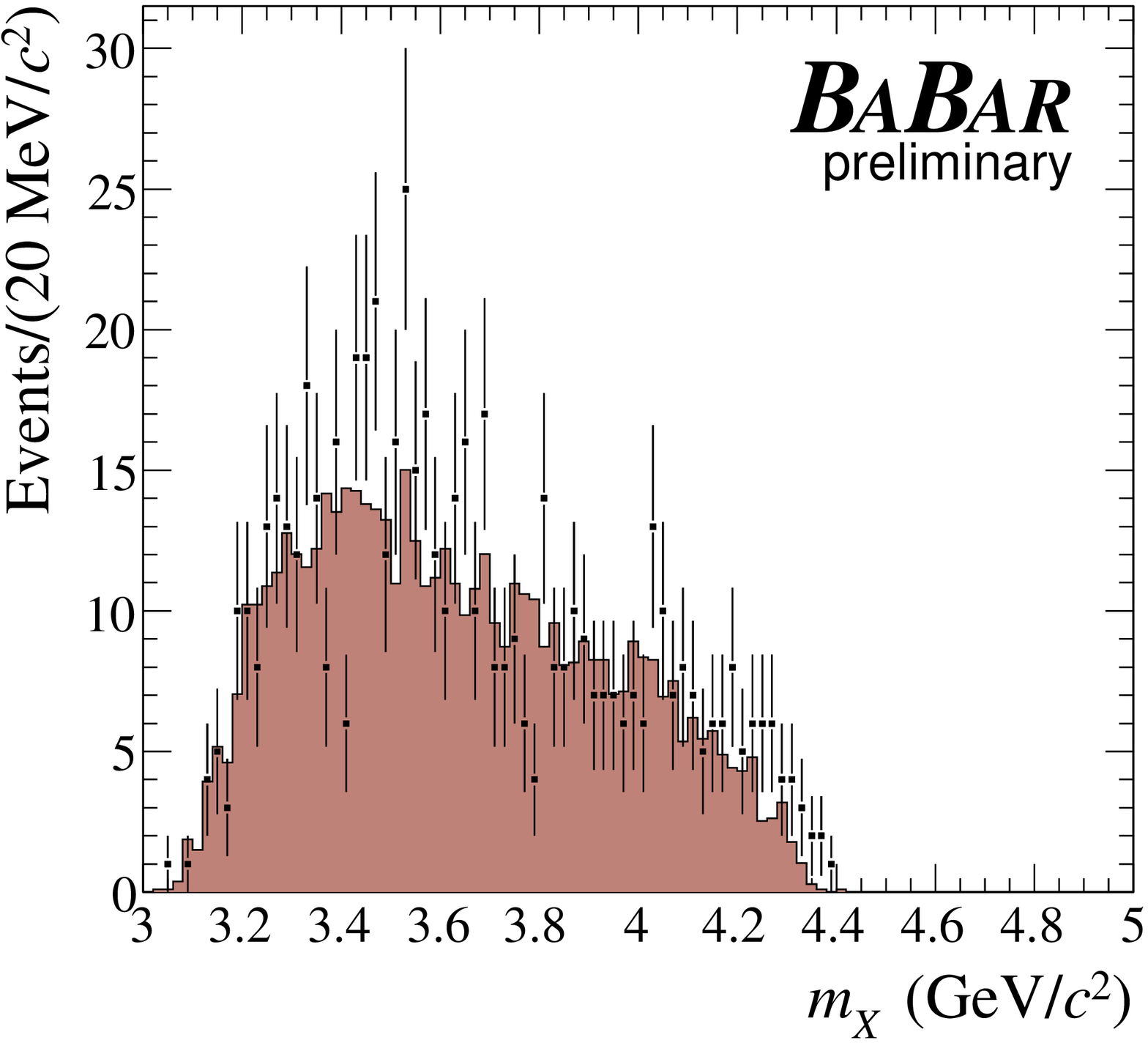}
\caption{Left: \mes\ distribution for $B^0\to (K\bar{K}\pi) K^{*0}$ (top), $\Bp\to\etac\gamma\Kp$ (middle) and $B^0\to\etac\gamma K^{*0}$ (bottom) candidates; points with error bars are data, the solid line represents the result of the fit described in the text, the dotted line represents the ARGUS component.
 Right: invariant mass distribution of the $K\bar{K}\pi (\gamma)$ system for $B^0\to (K\bar{K}\pi) K^{*0}$  (top), $\Bp\to\etac\gamma\Kp$ (middle) and $B^0\to\etac\gamma K^{*0}$ (bottom) decays; points with error bars are data in the \mes\ signal region, the shaded area represents the background expected from the \mes\ sideband. In the top plot, \etac\ and \jpsi\ peaks are visible; peaking background events are also present as signalled by the excess of the data points above the shaded area outside the two peaks. No appreciable $B$ component, neither signal nor peaking background, is observed for the $\Bp\to\etac\gamma\Kp$ and $B^0\to\etac\gamma K^{*0}$ cases.\label{fig:signalExtr_mes}}
  \end{center}
\end{figure}

We perform an unbinned maximum likelihood fit to the \mes\ distribution as follows. The \B\ component, accounting for the sum of signal and peaking background, is modelled by a Gaussian function whose width is taken from the simulation and whose mean is fixed to the $B$-meson mass~\cite{ref:pdg2006}. The \mes\ distribution of the combinatorial background is represented by an ARGUS threshold function~\cite{ref:Argus}. The total number of events and the ARGUS parameters are left free in the fit. The spectrum for candidates in the \mes\ sideband is normalized to the \mes\ signal window by using the integrals of the ARGUS component in the two regions (Fig.~\ref{fig:signalExtr_mes}). 

\begin{figure}[!htb]
  \begin{center} 
    \includegraphics[scale=0.4]{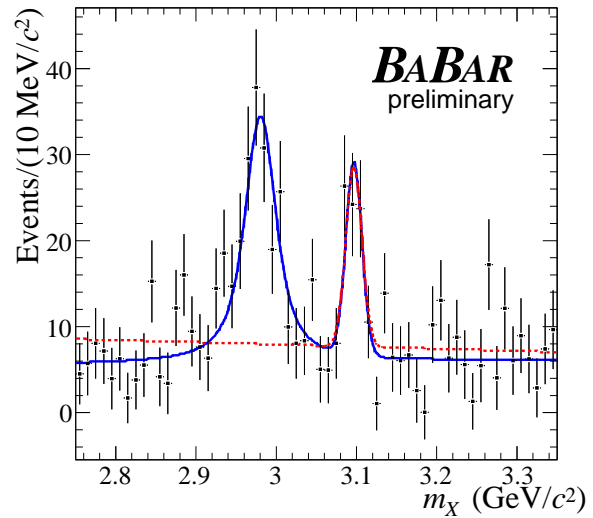}
\caption{\footnotesize{Fit result (solid blue line) superimposed on the \mes-sideband-subtracted \mX\ distribution (points with error bars) for $B^0\to (K\bar{K}\pi) K^{*0}$. The (red) dashed line is the result of the fit with no \etac\ component.}\label{fig:signalExtr_mx_etac}}
  \end{center}
\end{figure}

In the case of $B^0\to\etac \Kstarz$, the \mes-sideband-subtracted $m_X$ distribution is fitted to the sum of an \etac\ signal represented by a non-relativistic Breit-Wigner convolved with a Gaussian resolution function, a \jpsi\ component modelled by a Gaussian of the same width, and a background component accounted for by a first-order polynomial with free coefficients. The masses of the \etac\ and the \jpsi\ are fixed to the world average~\cite{ref:pdg2006}. The width of the \etac\ is fixed to the value measured by \babar, 34.3~\mevcc~\cite{ref:Babar_etac}. The mass resolution modelled by the width of the Gaussian functions is fixed to the value determined on MC events, $11~\mevcc$. The number of signal and background events is left free in the fit. We measure $185\pm22$ \etac\ and $59\pm12$ \jpsi\ candidates (Fig.~\ref{fig:signalExtr_mx_etac}). The $\chi^2$ value for the fit is 54 for 56 degrees of freedom ($N_{DoF}$). Repeating the fit with no \etac\ component we get a $\chi^2/N_{DoF}$ of $127/57$. A similar fit of the $\Bp\to\etac\Kp$ control sample yields $670\pm32$ \etac\ and $149\pm16$ \jpsi\ candidates with a $\chi^2/N_{DoF}$ of $86/56$. All of the values returned by these fits are summarized in Table~\ref{tab:yields_etac}.

\begin{table}[!htb]
\caption{Number of \etac\ ($N_{\etac}$) and \jpsi\ ($N_{\jpsi}$) events obtained from the fits described in the text with statistical errors only.}
\begin{ruledtabular}
\begin{tabular}{cccc} 
\rule[-1mm]{0mm}{1ex} Mode & $N_{\etac}$ & $N_{\jpsi}$ & $\chi^2/N_{DoF}$\\
\hline
\rule[-1mm]{0mm}{4ex} $B^0\to (\kkpi) \Kstarz$ & $185\pm22$      & $59\pm12$       & $54/56$ \\
\hline
\rule[-1mm]{0mm}{4ex} $B^+\to (\kkpi) K^+$     &  $670\pm32$      & $149\pm16$       & $86/56$ \\
\end{tabular}
\end{ruledtabular}
\label{tab:yields_etac}
\end{table}

\begin{figure}[!htb]
  \begin{center} 
    \includegraphics[scale=0.4]{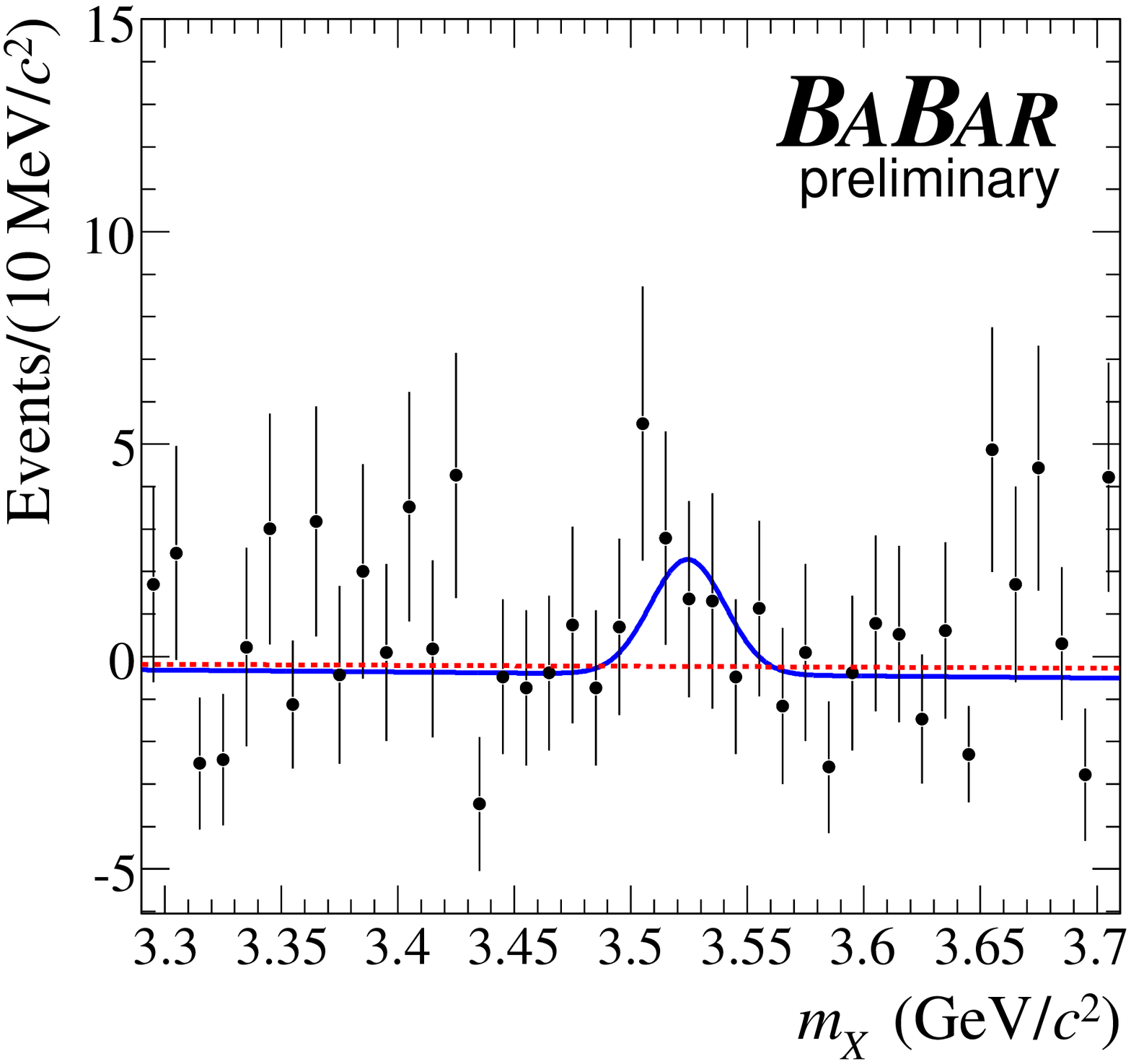}
    \includegraphics[scale=0.4]{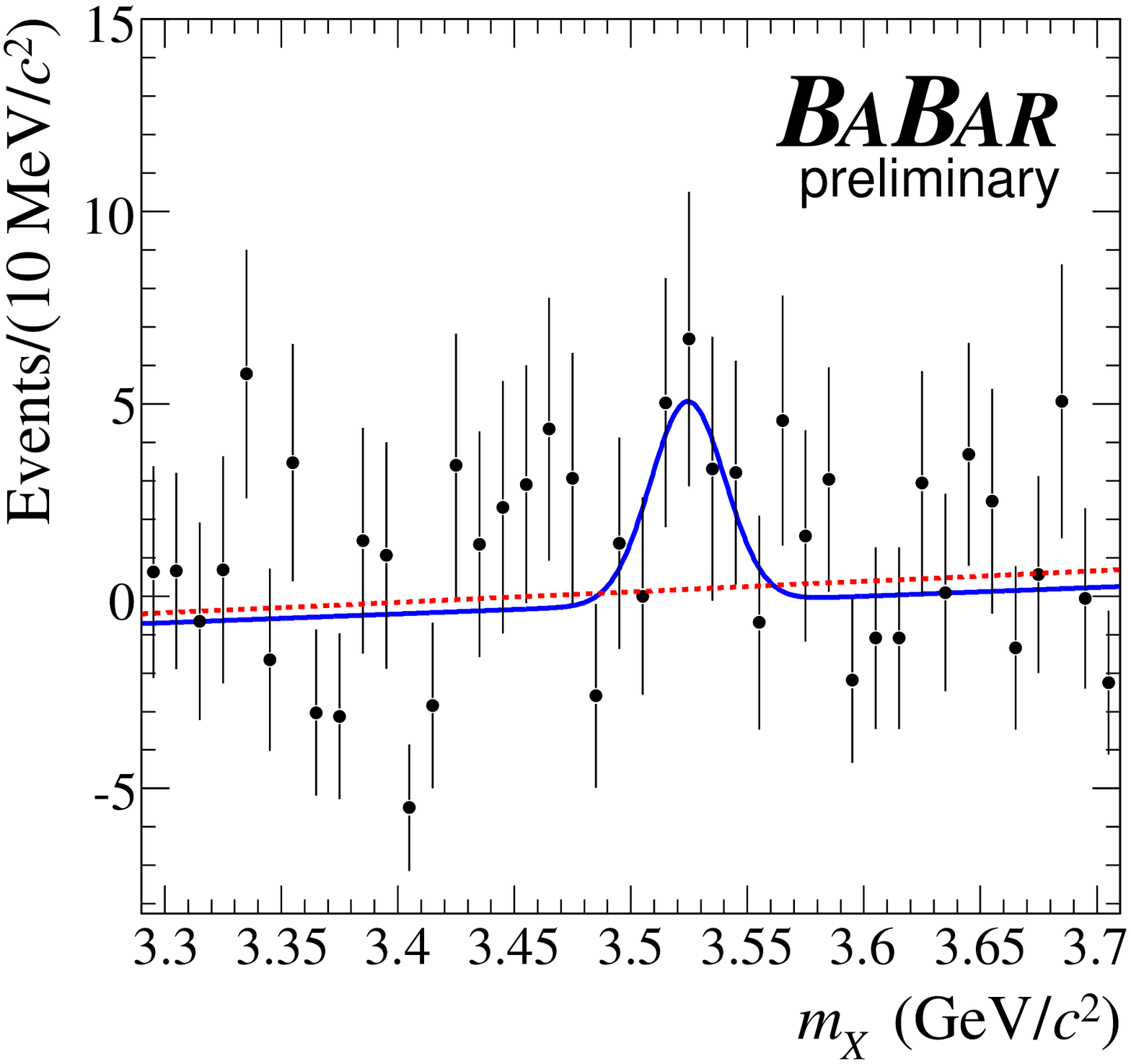}
\caption{\footnotesize{Fit result (solid blue line) superimposed on the \mes-sideband-subtracted \mX\ distribution (points with error bars) for $\Bp\to\etac\gamma\Kp$ (top) and $B^0\to\etac\gamma K^{*0}$ (bottom). No significant \hc\ signal is evident. The (red) dashed line is the result of the fit with no signal component.
}\label{fig:signalExtr_mx_hc}}
  \end{center}
\end{figure}

In the case of $\Bp\to\etac\gamma\Kp$ and $B^0\to\etac\gamma \Kstarz$, the \mes-sideband-subtracted $m_X$ distribution is fitted to the sum of  an \hc\ signal modelled by a Gaussian, and a background represented by a first-order polynomial. The mass of the \hc\ is fixed to the CLEO measurement, 3.524~\gevcc~\cite{ref:CLEO_1P1}. The Gaussian resolution is fixed to the value determined on MC events, $16~\mevcc$~\cite{ref:hc_simul}. In the fit, the number of signal and background events is left free. The fit is performed over the \mX\ range [3.3,3.7]~\gevcc. It yields $11\pm6$ and $21\pm8$ \hc\ candidates with a $\chi^2/N_{DoF}$ of $41/39$ and $42/39$ for the $B^+$ and $B^0$ yields, respectively (Fig.~\ref{fig:signalExtr_mx_hc} and Table~\ref{tab:yields_hc}). Repeating these fits with no \hc\ component, we get a $\chi^2/N_{DoF}$ of $45/40$ and $48/40$.

\begin{table}[!htb]
\caption{Number of \hc\ ($N_{\hc}$) obtained from the fits described in the text with statistical errors only.}
\begin{ruledtabular}
\begin{tabular}{ccc} 
\rule[-1mm]{0mm}{1ex} Mode & $N_{\hc}$ & $\chi^2/N_{DoF}$\\
\hline
\rule[-1mm]{0mm}{4ex} $B^+\to (\etac\gamma) K^+$     & $11\pm6$       & $41/39$ \\
\hline
\rule[-1mm]{0mm}{4ex} $B^0\to (\etac\gamma) \Kstarz$ & $21\pm8$       & $42/39$ \\
\end{tabular}
\end{ruledtabular}
\label{tab:yields_hc}
\end{table}
 
The stability of the fit results is verified for various configurations of the fitting conditions. We float the \etac\ mass and width, which are poorly known, and the \hc\ mass. The values for the signal yields and the floated parameters returned by these fits are consistent with the nominal configuration. We verify the goodness of the fit with the chosen model using a MC technique: we simulate a number of experiments by randomly generating samples of events distributed in \mX\ according to the models used in the fit. The number of events generated is equal to the number of events in the corresponding real data sample. The parameters of the distributions are set to their fixed or fitted values. The fit is repeated in the same conditions as on real data. The pulls for the number of signal and background events are distributed as expected. The robustness of the fit is tested on simulated events by varying the number of signal and background events input, including the null result. The number of events returned by the fit is consistent with the inputs for all cases. As an additional cross-check, we verify that the observed number of \jpsi\ candidates in the data agrees with the expectations.

We evaluate systematic uncertainties on the number of signal candidates by individually varying the parameters that are fixed in the fits by $\pm 1\sigma$ from their nominal values. We also estimate the systematic uncertainties that arise from a different choice of binning, fit range, and background parameterization. The large natural width of the \etac\ introduces the possibility of interference effects with non-resonant $B$ decays with the same final state particles that can modify the $m_X$ distribution with respect to the one used in the fit. The fit is repeated including an interference term between the \etac\ and the background in the fitting functions. The amplitude and phase of the interference term are left free in the fit. The variation of the \etac\ yield with respect to the nominal fit is taken as a conservative estimate of the systematic error due to neglecting interference effects. 
The total systematic uncertainty on the signal yield determination, summing in quadrature all the contributions, is $8.1$, $4.3$, $24.8$ and $18.1\%$ for $B^{+}\to\etac K^{+}$, $B^{0}\to\etac K^{*0}$, $B^{+}\to\hc K^{+}$ and $B^{0}\to\hc K^{*0}$, respectively. Being evaluated as yield variations on the data, most of these systematic uncertainties should improve with larger statistics.

The selection efficiency for $B^+\to\etac K^+$ is $6 \%$. The ratios of the selection efficiencies with respect to $B^+\to\etac K^+$, estimated by using simulated events, are $0.64\pm 0.01$, $0.51\pm 0.01$ and $0.29\pm0.02$ for $B^0\to\etac K^{*0}$, $B^+\to\hc K^{+}$ and $B^0\to\hc K^{*0}$, respectively. Most uncertainties on the efficiency cancel out in the ratios because of the similar final states. The remaining uncertainties are mainly due to differences between real data and simulation in the photon reconstruction as estimated from photon control samples on data ($1.8\%$), and the unknown polarization for $B^0\to\hc\Kstarz$ estimated as in~\cite{ref:Babar_chic2} ($6\%$). 

Using the signal efficiency computed on MC events, the signal yield observed on data, and the number of \BB\ pairs in the data sample, we derive $\BR(B^+\to\etac K^+)\times \BR(\etac\to K\bar{K}\pi)=(7.5\pm 0.4_{\mathrm{stat}})\times 10^{-5}$. This is in agreement with the world average $(6.4\pm 1.4)\times 10^{-5}$~\cite{ref:pdg2006}.

We calculate the ratios of the branching fractions with respect to $\BR(B^+\to\etac K^+)$ using the ratios of signal yields and efficiencies with respect to $B^+\to\etac K^+$, $R_{\Upsilon}=\Gamma(\FourS\to\B^+B^-)/\Gamma(\FourS\to B^0\bar{B}^0) = 1.026\pm 0.032$~\cite{ref:pdg2006}, and $\BR(\Kstarz\to K^+\pi^-)=2/3$, and summing the uncertainties in quadrature. Table~\ref{tab:systErr} summarizes the systematic uncertainties on the measurements. We obtain: 
\begin{equation}
\frac{\BR(B^0\to\etac K^{*0})}{\BR(B^+\to\etac\ K^+)} = 0.67 \pm 0.09_{\mathrm{stat}} \pm 0.07_{\mathrm{syst}}, \nonumber 
\end{equation}
and the $90\%$ C.L. upper limits
\begin{eqnarray}
\frac{\BR(B^+\to\hc K^+)\times \BR(\hc\to\etac\gamma)}{\BR(B^+\to\etac\ K^+)} &<& 0.058,\nonumber\\
\frac{\BR(B^0\to\hc K^{*0})\times \BR(\hc\to\etac\gamma)}{\BR(B^+\to\etac\ K^+)} &<& 0.26.\nonumber
\end{eqnarray}
These are determined assuming that the measurements follow a Gaussian distribution around the central value with standard deviation given by the total statistical plus systematic uncertainty.
\begin{table}[!hbt]
  \caption{Summary of the relative contributions to the systematic error on $R_{\etac K^*}=\BR(B^0\to\etac K^{*0})/\BR(B^+\to\etac\ K^+)$, $R_{\hc K}=\BR(B^+\to\hc K^+)\times \BR(\hc\to\etac\gamma)/\BR(B^+\to\etac\ K^+)$ and $R_{\hc K^*}=\BR(B^0\to\hc K^{*0})\times \BR(\hc\to\etac\gamma)/\BR(B^+\to\etac\ K^+)$.\\[0.5mm]}
  \begin{ruledtabular}
    \begin{tabular}{lccc} 
      & \multicolumn{3}{c}{$\sigma(R)/R$ $(\%)$}  \\
      \rule[-2mm]{0mm}{4ex}&   $R_{\etac K^*}$ & $R_{\hc K}$ & $R_{\hc K^*}$\\
\hline
Signal yield extraction                                    & 4.3 & 24.8 & 18.1\\
Signal efficiency                                          & 1.4 &  2.2 &  6.7\\
$\etac K^{+}$ yield extraction                             & 8.1 & 8.1 & 8.1       \\
$R_{\Upsilon}$                                                      & 3.1 & $-$    & 3.1\\ 
\hline
\rule[-1mm]{0mm}{4ex}Total                                   & 9.8 & 26.2 & 21.1\\
    \end{tabular}
  \end{ruledtabular}
  \label{tab:systErr}
\end{table}

Using $\BR(B^+\to\etac K^+) = (9.1\pm 1.3)\times 10^{-4}$, we derive 
\begin{equation}
\BR(B^0\to\etac K^{*0}) = (6.1\pm 0.8_{\mathrm{stat}} \pm 0.6_{\mathrm{syst}} \pm 0.9_{\mathrm{br}})\times 10^{-4},\nonumber
\end{equation}
where the last error is from the uncertainty on $\BR(B^+\to\etac K^+)$, and the $90\%$ C.L. upper limits
\begin{eqnarray}
\BR(B^+\to\hc K^+)\times \BR(\hc\to\etac\gamma) &<& 5.2\times 10^{-5},\nonumber \\
\BR(B^0\to\hc K^{*0})\times \BR(\hc\to\etac\gamma) &<& 2.41\times 10^{-4}.\nonumber 
\end{eqnarray}

Finally, we calculate 
\begin{equation}
\frac{\BR(B^0\to\hc K^{*0})\times \BR(\hc\to\etac\gamma)}{\BR(B^0\to\etac K^{*0})} < 0.39\nonumber
\end{equation}
at the $90\%$ C.L.

In summary, we obtain a measurement of $\BR(B^0\to\etac K^{*0})$ in agreement with, and greatly improving upon, the previous world average. We obtain an upper limit for $\BR(B^+\to\hc K^+)\times \BR(\hc\to\etac\gamma)$ in agreement with the result obtained by the previous Belle measurement, and a first upper limit on  $\BR(B^0\to\hc K^{*0})\times \BR(\hc\to\etac\gamma)$. The results confirm the suppression of \hc\ production in $B$ decays with respect to the S-wave \etac\ and the P-wave \chicone\ and \chiczero\ states. All results are preliminary.

\section{ACKNOWLEDGMENTS}
\label{sec:Acknowledgments}

We are grateful for the 
extraordinary contributions of our \pep2\ colleagues in
achieving the excellent luminosity and machine conditions
that have made this work possible.
The success of this project also relies critically on the 
expertise and dedication of the computing organizations that 
support \babar.
The collaborating institutions wish to thank 
SLAC for its support and the kind hospitality extended to them. 
This work is supported by the
US Department of Energy
and National Science Foundation, the
Natural Sciences and Engineering Research Council (Canada),
the Commissariat \`a l'Energie Atomique and
Institut National de Physique Nucl\'eaire et de Physique des Particules
(France), the
Bundesministerium f\"ur Bildung und Forschung and
Deutsche Forschungsgemeinschaft
(Germany), the
Istituto Nazionale di Fisica Nucleare (Italy),
the Foundation for Fundamental Research on Matter (The Netherlands),
the Research Council of Norway, the
Ministry of Science and Technology of the Russian Federation, 
Ministerio de Educaci\'on y Ciencia (Spain), and the
Science and Technology Facilities Council (United Kingdom).
Individuals have received support from 
the Marie-Curie IEF program (European Union) and
the A. P. Sloan Foundation.


\begin{thebibliography}{99}
\bibitem{ref:cc_note} Charge conjugated modes are implied throughout the paper.
\bibitem{ref:hc_simul} In the Monte Carlo events, the \hc\ meson is simulated with a $1~\mevcc$ total width.
\end{thebibliography}
\end{document}